\documentclass{article}
\usepackage[utf8]{inputenc}

\usepackage[hidelinks]{hyperref}
\usepackage{amsmath,amsfonts,amsthm,cleveref}
\usepackage[a4paper,top=2.9cm,bottom=2cm,left=2.5cm,right=2.5cm,marginparwidth=2cm]{geometry}
\usepackage{enumitem}
\usepackage{graphicx, subcaption}
\graphicspath{ {figures/} }
\usepackage{tikz}
\usetikzlibrary{decorations.pathreplacing}
\usepackage[sort&compress,numbers]{natbib}
\usepackage{todonotes}

\newcommand{\de}{\: \mathrm{d}}
\newcommand{\ddt}[2]{\frac{\mathrm{d}^2#1}{\mathrm{d}#2^2}}
\newcommand{\dd}[2]{\frac{\mathrm{d}#1}{\mathrm{d}#2}}
\newcommand{\ddp}[2]{\frac{\partial#1}{\partial#2}}
\newcommand{\xibf}{\hat{\xi}}
\newcommand{\loc}{\mathrm{loc}}
\newcommand{\fwhm}{\mathrm{FWHM}}

\newcommand{\sech}{\mathrm{sech}}
\newcommand{\Ai}{\mathrm{Ai}}
\newcommand{\I}{\mathrm{i}}
\newcommand{\argmax}{\mathrm{argmax}}
\newcommand{\R}{\mathbb{R}}

\numberwithin{equation}{section}

\title{On the problem of comparing graded metamaterials}
\author{Bryn Davies\thanks{Department of Mathematics, Imperial College London, London SW7~2AZ, UK (email: bryn.davies@imperial.ac.uk)} \and Lili Fehertoi-Nagy\thanks{Department of Physics, Imperial College London, London SW7 2AZ, UK} \and Henry J. Putley\footnotemark[1]}
\date{}

\begin{document}
\maketitle

\begin{abstract}
We use a simple effective model, obtained through the application of high-frequency homogenisation, to tackle the fundamental question of how the choice of gradient function affects the performance of a graded metamaterial. This approach provides a unified framework for comparing gradient functions efficiently and in a way that allows us to draw conclusions that apply to a range of different wave regimes. We consider the specific problem of single-frequency localisation, for which the appropriate effective model is a one-dimensional Schr\"odinger equation. Our analytic results both corroborate those of existing studies (which use either expensive full-field wave simulations or black-box numerical optimisation algorithms) and extend them to other metamaterial regimes. Based on our analysis, we are able to propose a design strategy for optimising monotonically graded metamaterials and offer an explanation for the lack of a universal optimal gradient function.
\end{abstract}


\section{Introduction} 

Structures with periodically modulated material parameters are ubiquitous in wave physics, thanks to their diverse and exotic wave propagation properties \cite{joannopoulos1995photonic, smith2004metamaterials, miltonbook}. Such materials appear in far-reaching applications, ranging from waveguides \cite{khanikaev2013photonic, rechtsman2013photonic} to perfect lenses \cite{pendry2000negative}, invisibility cloaks \cite{milton2006cloaking} and other metamaterials \cite{smith2004metamaterials, kadic20193d}. Fundamental to most of these fascinating applications is the existence of band gaps: ranges of frequencies that are unable to propagate through the material and for which the wave energy is reflected. Being able to predict and manipulate these band gaps has facilitated the development of many of the wave control devices now available.

\begin{figure}[t!]
    \centering
    \begin{tikzpicture}
    \foreach \x in {0,1,...,14}{
        \draw[fill=gray] (0.6*\x^0.85,0) circle (0.1);
    }
    \node at (0.3,0.5) {Low frequencies};
    \node at (4.5,0.5) {High frequencies};
    \node[white] at (6.7,-2) {.};

    \draw[->,thick] (-1,0) to (-0.5,0);

    \node at (0.5,1.5) {\textit{Monotonic gradient}};
    \end{tikzpicture}
    \includegraphics[width = 0.4\textwidth]{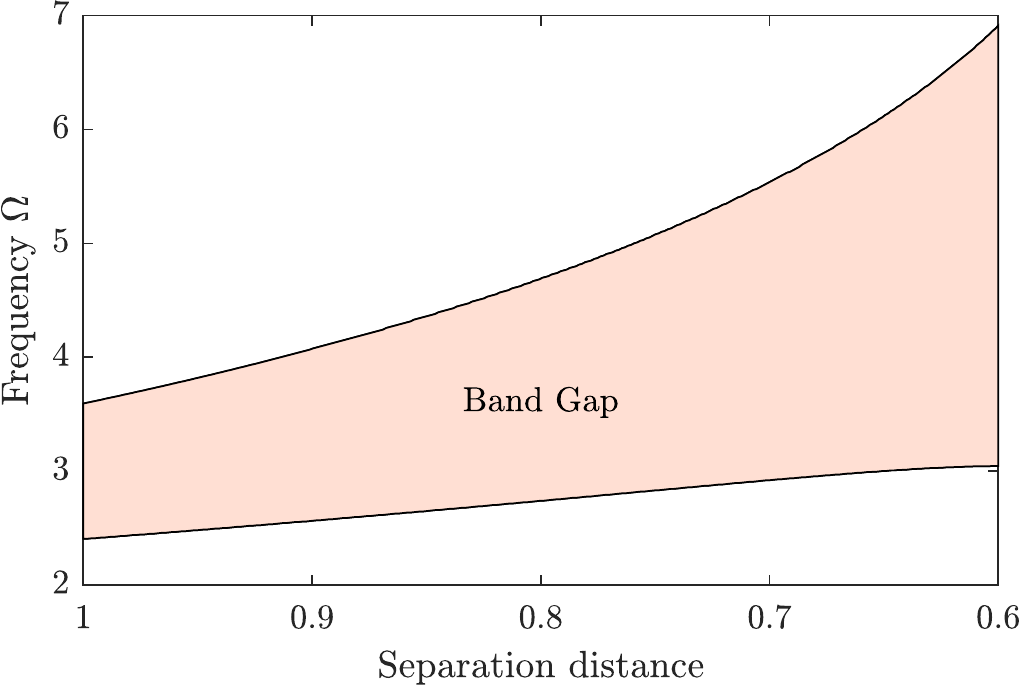}

    \vspace{0.5cm}
    
    \begin{tikzpicture}
    \foreach \x in {0,1,...,8}{
        \draw[fill=gray] (-0.55*\x^0.85,0) circle (0.1);
        \draw[fill=gray] (0.55*\x^0.85,0) circle (0.1);
    }
    \node[align=center] at (0,0.5) {Localisation};
    \node[white] at (4,-2) {.};

    \node at (-2,1.5) {\textit{Symmetric gradient}};
    \end{tikzpicture}
    \includegraphics[width = 0.4\textwidth]{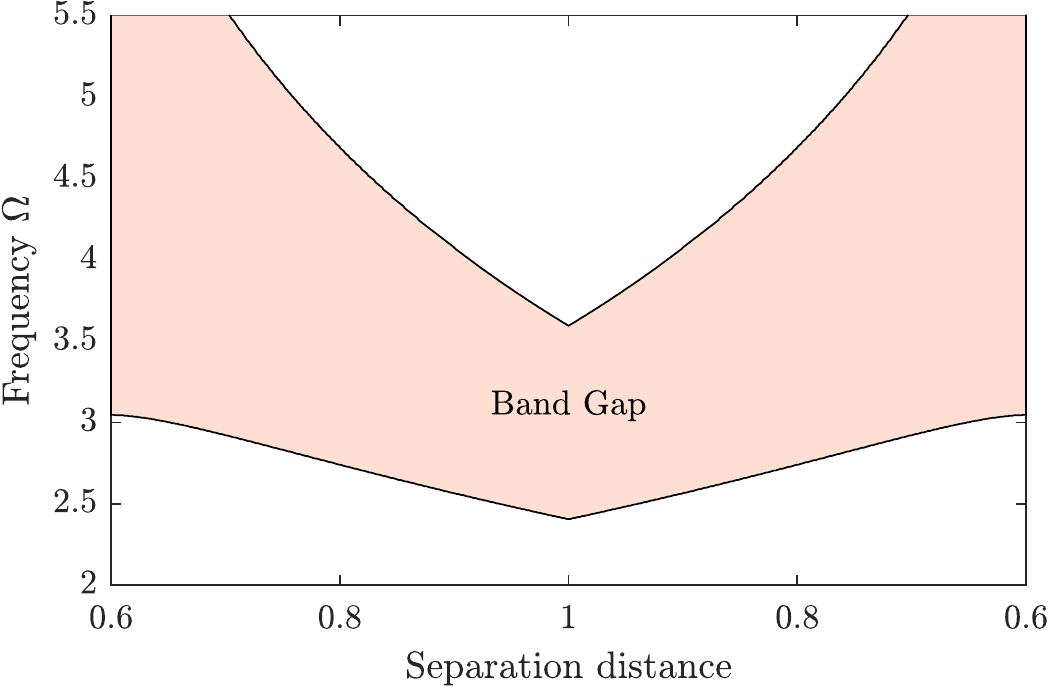}
    \caption{Slowly varying the separation distances between scatterers causes the local band gap to shift up or down. The top example is a monotonically graded metamaterial, where the separation distances gradually decrease causing the band gap to shift upwards. This configuration exhibits a rainbow effect. The lower example is a symmetrically graded metamaterial, where the separation distances increase before decreasing again. This configuration will lead to wave localisation at the centre of the array. These results are for arrays of cylinders with Neumann boundary conditions, modelled using multipole expansions. The cylinders have radius 0.25 and the horizontal axes in the plots show the separation distance between the cylinders' centres.}
    \label{fig:sketches}
\end{figure}

Graded metamaterials function by slowly varying the parameters of a band gap material in order to spatially modulate the local band gap. For example, varying geometric parameters such that the local band gap slowly shifts causes different frequencies to be reflected at different positions in the material. This phenomenon is often known as \emph{rainbow trapping} and was first demonstrated in an optical setting \cite{tsakmakidis2007trapped}. It has since been implemented in many other systems, including acoustics \cite{zhu2013acoustic}, elasticity \cite{arreola2019experimental, skelton2018multi}, seismic waves \cite{colombi2016seismic} and water waves \cite{bennetts2019low}. A simple example of such a system is shown in the upper part of \Cref{fig:sketches}, where we have modelled an array of cylinders with sound-hard (Neumann) boundary conditions. Varying the distances between the cylinders causes the local band gap to sweep upwards. This means that low frequencies will be reflected early on in the structure, while higher frequencies will be able to propagate further along the array before they encounter a local band gap. This simple phenomenon is referred to in this work as the \emph{rainbow effect} (to acknowledge the subtle differences between graded devices that ``trap'' or ``reflect'' energy, as discussed in \cite{chaplain2020delineating}).

The rainbow effect has found a range of significant applications, including energy harvesting and machine hearing. In the former case, the ability of a graded metamaterial to strongly focus waves of a specific frequency to a known location means a relatively small harvesting device (\emph{e.g.} a piezoelectric patch) can extract a relatively large amount of energy \cite{zhao2022graded, deponti2020graded}. Conversely, applications to machine hearing exploit the fact that a graded metamaterial exhibits a continuous relationship between the incoming frequency and the position of maximum amplitude \cite{ammari2020mimicking, marrocchio2021waves, rupin2019mimicking}. This replicates the action of the cochlea in mammalian hearing, and means that the frequency of a single-frequency wave can be identified by measuring the location at which the maximum amplitude occurs. 

If we allow the gradient of a graded metamaterial to not be monotonic, then we can produce other interesting phenomena. For example, if the gradient function is symmetric this causes the local band gap to shift in one direction and then back again. This is known to create localised eigenmodes in the centre of the array, as certain frequencies cannot propagate past a certain point in either direction \cite{antonakakis2014asymptotic}. The corresponding example for sound-hard cylinders is shown in \Cref{fig:sketches}. We see that the widened separation distances in the centre of the array mean that the band gap shifts downwards and then back up again. This means there is a range of frequencies (\emph{i.e.} around $\Omega=4$) that can propagate only in a central region of the array, so will be localised there.

A natural question to ask in the setting of graded metamaterials is what is the best gradient to choose? We need the spatial modulation to be on a slow scale, so that the physics of local band gaps applies, but otherwise are only limited by manufacturing constraints. As we will see in this work, this comparison is a somewhat subtle question. For this reason, many studies simply opt to use the obvious example of a linear gradient \cite{deponti2020graded, zhu2013acoustic, colombi2016seismic,chaplain2019rayleigh}. A notable exception to this is the field of cochlea-inspired graded metamaterials, where exponential gradients are the obvious choice to replicate the spatial frequency separation that occurs in the cochlea \cite{ammari2019fully, rupin2019mimicking, marrocchio2021waves}.

There have been some previous studies comparing the effect of different gradients on metamaterial function. For example, \cite{rg_linear, rg_exponential} compare the localisation of sound waves in graded acoustic metamaterials and make the (qualitative) observation that an exponentially graded crystal yields ``a stronger wave enhancement'' than a linear gradient. Conversely, \cite{zhao2022graded} observes that in a flexural system of graded beam-like resonators a cube-root gradient yields a ``stronger bandgap'' and ``higher energy density'' than both linear and cubic gradients. Similarly, \cite{alshaqaq2020graded} shows that the ``largest power is harvested'' by a square-root gradient, compared to a linear or quadratic gradient. These examples demonstrate the lack of consensus in the field, particularly since the gradients they settle on are so different (the exponential gradient is convex whereas the square and cube roots are concave, for example). This motivates the present study, where we develop a concise approach for systematically characterising the localisation due to different gradient functions.

A natural approach when seeking to optimise the performance of a graded metamaterial is to use machine learning and numerical optimisation. For example, \cite{jimenez2017rainbow} used constrained optimisation to maximise the broadband absorption of a graded array of Helmholtz resonators and \cite{wilks2022rainbow} extended this approach to study water waves trapped by a graded array of vertical barriers. Similarly, \cite{rosafalco2022optimised} used reinforcement learning to optimise a graded array of resonators on an elastic plate. It is worth noting that these sophisticated numerical studies often lead to simple answers: for example, the final configuration obtained by \cite{jimenez2017rainbow} for a broadband rainbow absorber is an array of Helmholtz resonators whose resonant frequencies vary (approximately) linearly as a function of their position. In this work, we will favour an approach that reveals the underlying mechanisms to explain the relative performance of different gradient functions. Further, the effective model we consider means we are able to draw conclusions that transcend the specific physical setting.

There have been some other recent breakthroughs in the field of graded metamaterials that should be mentioned, although they are not central to this study. While subwavelength graded metamaterials inherently have some robustness to imperfections \cite{davies2022robustness}, the principle of ``topological protection'' has been used to create graded structures with enhanced robustness properties \cite{chaplain2020topological}. Further, it has been shown that using periodic materials with zero group velocity modes \emph{within} the first Brillouin zone leads to graded structures that genuinely ``trap'' the energy, rather than just reflecting it, increasing the harvested energy \cite{chaplain2020delineating}. Additionally, it has been shown that quasicrystalline arrangements of scatterers can be graded to produce devices that display a rainbow effect with a fractal collection of many band gaps, giving a very broadband response \cite{davies2023graded}.

A central question underpinning this study is to decide on a quantitative measure of wave localisation in graded media. Indeed, a possible explanation for the lack of consensus over the optimal gradient function is that there are different ideas on the quantity that should be optimised. Suggestions include maximising the energy density at a specific location \cite{zhao2022graded, wilks2022rainbow} or maximising the total absorption over a range of frequencies \cite{wilks2022rainbow, jimenez2017rainbow}. In the present study, we will consider the problem of maximising the focusing of a single-frequency wave. In doing so, we are optimising the extent to which energy is localised, ready to be harvested. Similarly, stronger localisation increases the resolution of any frequency sensor based on a graded metamaterial. Hence, this motivates the problem of studying the profile of the localised eigenmodes, measured by normalising the amplitude and measuring the width using the full width at half maximum.

We will take advantage of a multi-scale approximation method known as \emph{high-frequency homogenisation} \cite{craster2010high} to study the profile of localised eigenmodes. This is a two-scale approach, similar to the work of \emph{e.g.} \cite{allaire1992homogenization, bensoussan2011asymptotic}, that can be used to model graded materials as perturbations of the original periodic materials \cite{antonakakis2014asymptotic, craster2022asymptotic}. It does so by characterising the localised eigenmodes as being, at leading order in the asymptotic parameter, the product of a slowly varying amplitude function (sometimes referred to as the \emph{envelope}) and the Bloch mode that exists in the unperturbed periodic structure. It is already well established that the homogenised model gives an accurate prediction of a localised eigenmode's amplitude; see \emph{e.g.}  \cite{craster2010high, antonakakis2014asymptotic, craster2022asymptotic} and references therein, as well as the convergence analyses of \cite{guzina2019rational, meng2022convergent}. This approach means we can make meaningful comparisons between different gradient functions without the need for any direct numerical studies of the wave scattering problem. Additionally, it means that our conclusions apply to a wide range of physical systems, as several different wave scattering problems have been shown to reduce to the same effective model.

This article will begin by summarising the key steps in the derivation of the effective model through high-frequency homogenisation in Section~\ref{sec:homogenisation}, before moving on to studying solutions of the resulting ordinary differential equation. After stating the precise formulation of the localisation problem we will consider in \Cref{sec:localisation}, including the appropriate reformulation of the full width at half maximum, we will move on to studying symmetric gradients in \Cref{sec:symmetric} and monotonic gradients in \Cref{sec:monotonic}. In both cases, we will present both numerical solutions (computed using spectral differentiation) and analytic solutions (subject to appropriate approximations, in some cases).

\section{Effective model} \label{sec:homogenisation}

The analysis in this paper will all be concerned with understanding the localised solutions to an effective model. This model is a one-dimensional Schr\"odinger equation that characterises wave propagation in perturbed periodic media. Given a gradient function $g:\mathbb{R}\to\mathbb{R}$ and parameters $T,\alpha>0$, which depend on the unperturbed periodic material, we consider solutions $f$ to the eigenvalue problem
\begin{equation} \label{eq:ODE}
	-T f''(X)+\alpha g(X) f(X)=\Omega_2^2 f(X).
\end{equation}
The eigenvalue $\Omega_2^2$ corresponds to the difference between the eigenfrequency of the localised eigenmode and that of the Bloch mode at the cutoff frequency of the unperturbed periodic structure.

The Schr\"odinger model \eqref{eq:ODE} arises naturally in several settings, such as through the application of high-frequency homogenisation to modelling localised eigenmodes in perturbed periodic media \cite{craster2010high}. We will present two such examples of this below. The first is for a one-dimensional periodic material that is graded by scaling the wave speed according to the function $g$. Since this analysis is strictly novel (but similar to previous works) we will briefly present the details of the derivation. In two dimensions, similar analyses were conducted in \cite{antonakakis2014asymptotic}, where the gradient is applied as a re-scaling of the material geometry, in the direction perpendicular to the axis of periodicity. This was shown to apply for a range of physical systems, including a diffraction grating composed of cylinders and a metasurface with protruding comb-like teeth, as sketched in Figure~\ref{fig:gradients}. We will summarise the key assumptions of their analysis here, for completeness. In both cases, the interpretation of \eqref{eq:ODE} is that $f$ is the slowly varying amplitude of the eigenmode, $T$ is the effective group velocity and $\alpha$ is a parameter that describes (in conjunction with the eigenvalue $\Omega_2^2$) how the eigenfrequency of the localised eigenmode is perturbed away from the edge of the band gap.

Both the two simple examples we outline below in Sections~\ref{sec:1dHFH} and~\ref{sec:2dHFH} have material parameters that are non-dispersive (the local wave speed $c$ does not depend on the frequency $\Omega$). However, high-frequency homogenisation can also be applied to locally dispersive materials \cite{touboul2023dispersive}. In this case, the resulting effective equations for $f$ are the same, with the dispersive material parameters appearing in the expressions for the constants $T$ and $\alpha$. This means that the effective model \eqref{eq:ODE} can also be used to describe systems with local resonances, as is the case for several of the graded metamaterials explored in the literature, \emph{e.g.} \cite{ammari2020mimicking, colombi2016seismic, deponti2020graded, marrocchio2021waves}. The information about the local resonances is contained within the expressions for $T$ and $\alpha$.

\subsection{One-dimensional analysis} \label{sec:1dHFH}

We study a one-dimensional system, which describes the propagation of waves through an alternating sequence of materials
\begin{equation} \label{eq:helmholtz1d}
l^2 \ddt{u}{x} + \frac{\Omega^2}{c^2} u=0,
\end{equation}
where $\Omega$ is the non-dimensionalised frequency and $l$ is the length of the periodicity of the problem. If $l$ is the ``short'' scale, then we let $L$ be the ``long'' scale, which characterises the overall length of the structure. We suppose that $\epsilon=l/L\ll1$ and assume that the wave speed $c=c(x/l)$ varies as a function of the short-scale variable.

We suppose that the material is graded according to some function $g$, in the sense that the material parameters are scaled by a varying factor $1-\epsilon^2 g(X)$. If we define the coordinates 
\begin{equation}
\xi=\frac{x}{l}, \quad X=\frac{x}{L},
\end{equation}
then we suppose that the distribution of the material parameters $c$ is graded according to
\begin{equation}
    c(\xi)\mapsto\left( 1-\epsilon^2 g(X) \right)^{-1} {c}(\xi).
\end{equation} 
Under this transformation, the equation \eqref{eq:helmholtz1d} becomes
\begin{equation} \label{eq:rescaledH}
    \ddp{^2u}{\xi^2} + 2\epsilon\ddp{^2u}{X \partial\xi} + \epsilon^2 \ddp{^2u}{X^2} + \left( 1-2\epsilon^2g(X) + \epsilon^4g(X)^2 \right)\frac{\Omega^2}{{c}^2(\xi)} u=0.
\end{equation}
We subsequently seek a multi-scale solution of the form
\begin{align} \label{eq:multiscale1d}
u(\xibf,X) &= u_0(\xibf,X)+\epsilon u_1(\xi,X) +\epsilon^2 u_2(\xi,X)+\dots,
\end{align}
where the key step in the high-frequency homogenisation methodology \cite{craster2010high} is to make assumptions on the leading-order terms. In particular, we are interested in the localisation of waves that occurs at the edges of a band gap as that gap shifts due to the $O(\epsilon^2)$ perturbations. Thus, we assume that $\Omega$ is a perturbation of $\Omega_0$, which is the cutoff frequency at the edge of the band gap:
\begin{align}
\Omega^2&=\Omega_0^2+\epsilon \Omega_1^2 + \epsilon^2\Omega_2^2+\dots.
\end{align}
We assume that $\Omega_0^2$ is a simple eigenvalue, which is not a restrictive assumption since we want it to be at the edge of a band gap. Seeking a solution of the form \eqref{eq:multiscale1d}, we find that the leading-order equation in $\epsilon$ is
\begin{equation} \label{eq:eps0}
    \ddp{^2u_0}{\xi^2}+\frac{\Omega_0^2}{c^2(\xi)}u_0=0,
\end{equation}
from which we see that 
\begin{equation} \label{eq:U0ansatz}
u_0(\xi,X) = U_0(\xi)f(X),
\end{equation}
where $U_0$ is the standing wave that exists at the cutoff frequency and $f(X)$ is a slowly varying function that describes the envelope of the mode. 

Moving to the $O(\epsilon)$ terms in the hierarchy of equations, we find that 
\begin{equation} \label{eq:eps1}
    \ddp{^2u_1}{\xi^2}+\frac{\Omega_0^2}{c^2(\xi)}u_1=-2\ddp{^2u_0}{\xi \partial X}-\frac{\Omega_1^2}{c^2(\xi)} u_0.
\end{equation}
If \eqref{eq:eps1} is multiplied by $u_0$ and integrated for $\xi$ in the periodic unit cell $\xi\in[-1,1]$, then the periodic or anti-periodic boundary conditions at either end imply that $\Omega_1=0$. Thus, we conclude that $u_1$ can be written as $u_1(\xi,X)=f'(X)U_1(\xi)+h(X)U_0(\xi)$ where $f'$ is the derivative of the envelope function $f(X)$ defined in \eqref{eq:U0ansatz}, $h$ is some unknown function and $U_1$ satisfies the equation
\begin{equation}
    \ddt{U_1}{\xi}+\frac{\Omega_0^2}{c^2(\xi)}U_1=-2\dd{U_0}{\xi}.
\end{equation}
This can be solved to find $U_1$ explicitly\footnote{In particular, if $W(\xi)$ is any non-periodic solution of the leading-order equation \eqref{eq:eps0}, then we have that $U_1(\xi)=AW(\xi)-\xi U_0(\xi)$, where $A$ is a constant given by $A=2U_0(1)/(W(1)+W(-1))$. See \cite{craster2010high} for details.}. 

Finally, we turn to the $O(\epsilon^2)$ terms in \eqref{eq:rescaledH}. The corresponding equation is
\begin{equation} \label{eq:eps2}
    \ddp{^2u_2}{\xi^2}+\frac{\Omega_0^2}{c^2(\xi)}u_2= 2g(X) \frac{\Omega_0^2}{c^2(\xi)}u_0 -2 \ddp{^2 u_1}{\xi \partial X} -\ddp{^2u_0}{X^2} -\frac{\Omega_2^2}{c^2(\xi)}u_0.
\end{equation}
Assuming the periodicity in $c(\xi)$ is with respect to the unit cell $\xi\in[-1,1]$, we can multiply \eqref{eq:eps2} by $U_0(\xi)$ and integrate over $\xi\in[-1,1]$ to find that
% \begin{equation}
%     \int_{-1}^1 2g(X) \frac{\Omega_0^2}{c^2(\xi)}u_0U_0 +2 \ddp{^2 u_1}{\xi \partial X}U_0 +\ddp{^2u_0}{X^2}U_0 +\frac{\Omega_2^2}{c^2(\xi)}f(X)U_0(\xi)^2 \de\xi =0,
% \end{equation}
\begin{equation}
    \int_{-1}^1 2 \frac{\Omega_0^2}{c^2(\xi)}g(X)f(X)U_0(\xi)^2 -2f''(X)U_1'(\xi)U_0(\xi) -f''(X)U_0(\xi)^2 -\frac{\Omega_2^2}{c^2(\xi)}f(X)U_0(\xi)^2 \de\xi =0.
\end{equation}
Rearranging the terms, we see that $f$ satisfies the eigenvalue problem \eqref{eq:ODE} with $T$ given by
\begin{equation} \label{eq:T1D}
    T=\frac{\int_{-1}^1 2U_1'(\xi)U_0(\xi)+U_0^2(\xi) \de\xi}{\int_{-1}^1 U_0^2(\xi)/c^2(\xi)\de\xi}
\end{equation}
and $\alpha=2\Omega_0^2$.

\begin{figure}
    \centering
    \includegraphics[width=0.7\linewidth]{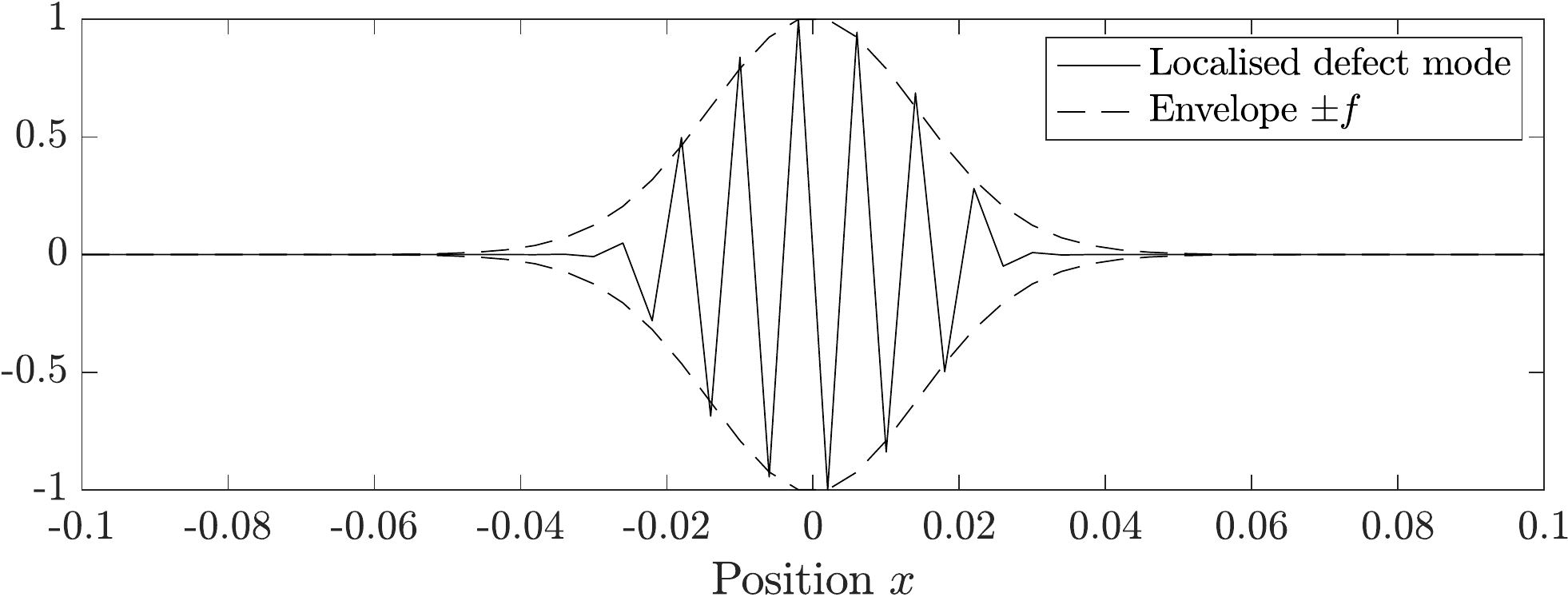}
    \caption{The localised defect mode for a one-dimensional two-phase periodic material that is perturbed by the symmetric gradient $g(X)=1-\mathrm{sech}^2X$. In this case, $\epsilon=0.05$ and the homogenised envelope function, obtained by solving \eqref{eq:ODE}, accurately predicts the decay of the amplitude. }
    \label{fig:sechmode}
\end{figure}

An example of a localised mode and the accompanying homogenised envelope $f(X)$ are shown in \Cref{fig:sechmode}, for the case where $g(X)=1-\mathrm{sech}^2X$. We can see that the envelope, which was obtained by solving \eqref{eq:ODE} with the appropriate values for $T$ and $\alpha$, predicts the rate of decay of the localised defect mode well. In this case, the asymptotic parameter has the value $\epsilon=0.05$. In other works, this method has been shown to be (surprisingly) robust with respect to the choice of $\epsilon$ and gives a good approximation even when it is comparatively large, see \emph{e.g.} \cite{antonakakis2014asymptotic, craster2022asymptotic}.

\subsection{Two-dimensional analysis} \label{sec:2dHFH}

\begin{figure}
    \centering
    \includegraphics[width = 0.9\textwidth]{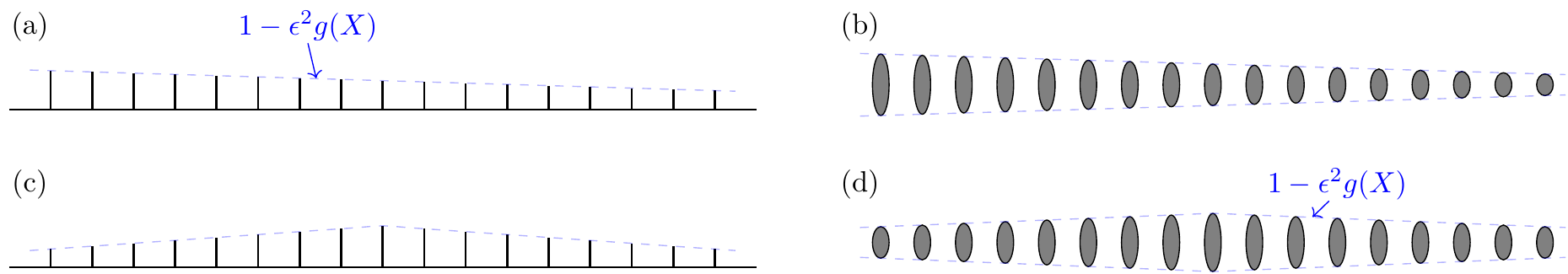}
    \caption{High-frequency homogenisation can be used to derive the effective equation that is the basis of this work from a range of different graded wave scattering problems. For example, the two-dimensional analysis presented in Section~\ref{sec:2dHFH} has been shown to apply both to a metasurface with comb-like teeth, as depicted in (a) and (c), and to a diffraction grating composed of elliptical cylinders, as depicted in (b) and (d). We show examples of both monotonic and symmetric, with the gradient function indicated with a dashed line in each case.}
    \label{fig:gradients}
\end{figure}

In two-dimensions, we consider time-harmonic waves propagating through a homogeneous medium that occupies the exterior of some scattering material. This material could be, for example, a metasurface with protruding comb-like teeth or a diffraction grating composed of elliptical cylinders, as sketched in Figure~\ref{fig:gradients}. This model gives a good description of the scattering of transverse electric (TE) polarised electromagnetic waves by a perfectly conducting material if we apply Neumann boundary conditions on the surface of the scattering material. 

We suppose that the unperturbed material is periodic in the direction parallel to the $x_1$ axis. The governing equation is 
\begin{equation}
l^2 \Delta u + \Omega^2 u=0,
\end{equation}
where $\Omega$ is the non-dimensionalised frequency. Once again, we let $l$ be the ``short'' scale and $L$ be the ``long'' scale, supposing that $\epsilon=l/L\ll1$. In this case, the gradient will be applied by rescaling the geometry in the direction perpendicular to the direction of periodicity. Again, this will be done by multiplying by a factor $1-\epsilon^2 g$ (this time, in the $x_2$-direction). We can define the new coordinates 
\begin{equation}
\xi_1 = \frac{x_1}{l}, \quad \xi_2=\frac{x_2}{l}\left(1-\epsilon^2g(X)\right), \quad X=\frac{x_1}{L},
\end{equation}
and subsequently seek a multi-scale solution to \eqref{eq:ODE} of the form
\begin{align}
u(\xibf,X) = u_0(\xibf,X)+\epsilon u_1(\xibf,X) +\epsilon^2 u_2(\xibf,X)+\dots,
\end{align}
where $\xibf=(\xi_1,\xi_2)$. As in the one-dimensional case, we suppose that $\Omega^2=\Omega_0^2+\epsilon \Omega_1^2 + \epsilon^2\Omega_2^2+\dots$, where $\Omega_0$ is the cutoff frequency, corresponding to a simple eigenvalue whose eigenmode takes the form of a Bloch mode that is either periodic or anti-periodic. In this case, solving the leading-order equation in $\epsilon$ once again tells us that 
\begin{equation}
u_0(\xibf,X) = U_0(\xibf)f(X),
\end{equation}
where $U_0$ is the standing wave that exists at the cutoff frequency and $f(X)$ is a slowly varying function that describes the envelope of the mode. Proceeding to solve the hierarchy of equations in $\epsilon$, we find that $\Omega_1=0$ and that $f$ and $\Omega_2$ must satisfy the Schr\"odinger equation \eqref{eq:ODE}, where $T$ and $\alpha$ are parameters which capture the microstructure of the unperturbed periodic problem, given by
\begin{equation}
     T = \dfrac{-\int \int_{S} (U_0^2 + U_0\ddp{}{\xi_1}U_{1} - U_1\ddp{}{\xi_1}U_{0}){\rm d}S}{\int \int_S U_0^2 {\rm d}S}
\end{equation}
and
\begin{equation}
    \alpha = \dfrac{\int \int_S (\ddp{}{\xi_2}U_{0}^2 - U_0 \ddp{^2}{\xi_2^2}U_{0}){\rm d}S}{\int \int_S U_0^2 {\rm d}S},
\end{equation}
where $U_0=U_0(\xi_1,\xi_2)$ is the standing wave that exists at the cutoff frequency $\Omega_0$, $S$ is the exterior of the scattering material within unit cell and $U_1(\xi_1,\xi_2)$ is the solution to the inhomogeneous Helmholtz problem
\begin{equation}
\Delta_{\xibf} U_1 + \Omega_0^2 U_1 = - 2\partial_{\xi_1} U_0,
\end{equation}
with the appropriate periodic or anti-periodic boundary conditions. See \cite{antonakakis2014asymptotic} for details of this derivation.

\subsection{Admissible parameter values}

In the rest of this paper, we will focus solely on solving the Schr\"odinger equation \eqref{eq:ODE} that we have motivated through high-frequency homogenisation. It has been shown repeatedly in previous works that this model gives good approximations to the one- and two-dimensional physical problems described above, even when $\epsilon$ is relatively large. An example is shown in Figure~\ref{fig:sechmode}, where a one-dimensional periodic material is perturbed with the gradient function $g(X)=1-\sech^2(X)$ to create a localised mode. Many further examples can also be found in \emph{e.g.} \cite{craster2010high, guzina2019rational, meng2022convergent, craster2022asymptotic, assier2020high, antonakakis2014asymptotic} and references therein.

Before we start solving the homogensised Schr\"odinger equation \eqref{eq:ODE} for different gradient functions $g(X)$, it is important to understand the range of values of the parameters $T$ and $\alpha$ that are relevant. As a demonstrative example, consider the simple one-dimensional problem \eqref{eq:helmholtz1d} with wave speed given by
\begin{equation} \label{eq:1dmaterial}
    c(\xi)=\begin{cases}
    1 & \text{for } -1\leq\xi<0, \\
    1/r & \text{for } 0\leq\xi<1,
    \end{cases}
\end{equation}
where $r\neq1$ is some real-valued parameter. The dispersion relation relating the frequency $\Omega$ and the Bloch parameter $\kappa$ is well known \cite{kronig1931quantum}:
\begin{equation}
    \cos\Omega\cos\Omega r -\frac{1+r^2}{2r}\sin\Omega\sin\Omega r=\cos2\kappa.
\end{equation}
This can be solved when $\kappa=0,\pi$ to find the appropriate cutoff frequencies $\Omega_0$. This example was studied in detail in \cite[Section 3a]{craster2010high} where it was calculated that the formula \eqref{eq:T1D} yields the expression
\begin{equation} \label{eq:T1Dexample}
    T=\frac{\pm4\Omega_0\sin\Omega_0\sin\Omega_0r}{(r\sin\Omega_0\mp \sin\Omega_0r)(\cos\Omega_0\mp\cos\Omega_0r)},
\end{equation}
where the upper/lower signs should be chosen depending on whether the Bloch mode at the cutoff frequency $\Omega_0$ is periodic/anti-periodic.

\begin{figure}
    \centering
    \includegraphics[width=0.8\linewidth]{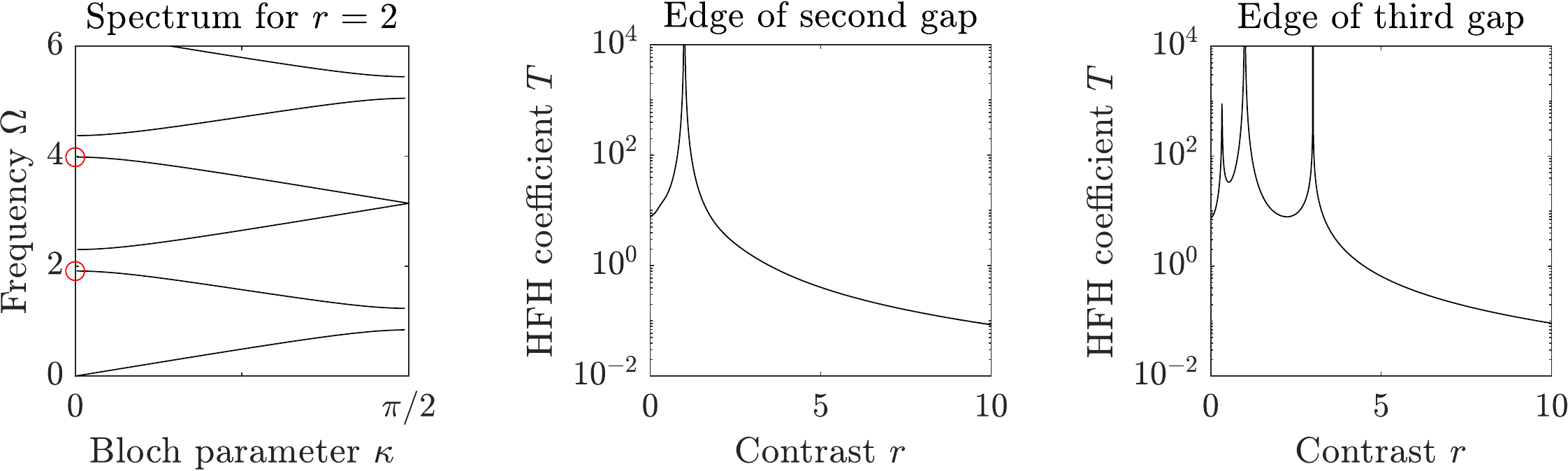}
    \caption{For a simple example of a one-dimensional two-phase material, the dispersion relation and the homogenised coefficients can be computed explicitly. Here, we show the Bloch spectral bands for the material \eqref{eq:1dmaterial} in the specific case when the material contrast parameter $r=2$. Then, we show the high-frequency homogenisation coefficient $T$, given by \eqref{eq:T1Dexample}, as a function of $r$, for $\Omega_0$ being at the lower edge of the second and third band gaps, respectively.}
    \label{fig:Texample}
\end{figure}

In general, when $\Omega_0$ is at the lower edge of a band gap, it will be the case that $T>0$. Some examples of $T$ (plotted as a function of the material contrast parameter $r$) are shown in Figure~\ref{fig:Texample}, corresponding to the cutoff frequencies at the lower edge of the second and third band gaps. An example of the Bloch spectrum is given in the left-hand plot of Figure~\ref{fig:Texample}, for $r=2$, where the corresponding cutoff frequencies are indicated by red circles. We can see that $T$ is singular for some values of $r$ (including $r=1$, for example). These are the values at which the corresponding band gap closes. This violates our assumption that $\Omega_0$ is a simple root and means the effective model must be modified. This can be done (we typically no longer have $\Omega_1=0$, see \emph{e.g.} \cite{craster2010high}), however the model for simple eigenvalues is sufficient for our purposes, since we are interested in eigenvalues $\Omega_0$ that lie at the edges of band gaps.

The main conclusion to draw from Figure~\ref{fig:Texample} for this study is that $T$ ranges over several orders of positive values, including taking small values when $r$ is large. If we instead took examples of $\Omega_0$ which are at the upper edge of a band gap, then $T$ would take negative values. However, in this case we would also have that $\Omega_2^2<0$ and would need to make the change $g\mapsto-g$ to achieve the same rainbow effect. Hence, it is sufficient to study \eqref{eq:ODE} with $T,\alpha>0$ and we need to consider $T/\alpha$ ranging over several orders of magnitude.

\section{Localisation problem statement} \label{sec:localisation}

The specific problem we will consider in this study is how the choice of gradient function affects the ability of a graded metamaterial to localise a wave of a single frequency. We will use the canonical effective model \eqref{eq:ODE} as a unified framework for comparing metamaterials with different gradient functions $g$, independent of the specific physical setting. That is, given a gradient function $g(X)$, we will explore the extent to which the envelope function $f(X)$, obtained by solving \eqref{eq:ODE}, is localised. 

To quantify the localisation of the slowly varying envelope function $f(X)$, we will use a version of the \emph{full width at half maximum} (FWHM). This measures the width of a function by taking the distance between the furthest two points at which the function is equal to at least half of its maximum value. Our version will be modified such that we only measure the width of the largest peak. Then, assuming the function $f$ is continuous and has a unique $x_\mathrm{max}:=\argmax_{x\in\R}|f(x)|$ we define the FWHM as 
\begin{equation} \label{eq:fwhm_defn}
    \fwhm(f) = \inf\big\{x>x_\mathrm{max}:|f(x)|<\tfrac{1}{2}|f(x_\mathrm{max})|\big\}-\sup\big\{x<x_\mathrm{max}:|f(x)|<\tfrac{1}{2}|f(x_\mathrm{max})|\big\}.
\end{equation}
Then, given a metamaterial with high-frequency homogenisation coefficients $T$ and $\alpha$, our quantitative measure of wave localisation in a graded metamaterial will be 
\begin{equation} \label{def:loc}
    \loc(g)_{T,\alpha} = \frac{1}{\fwhm(f)},
\end{equation}
where the corresponding $f$ is the first eigenmode of \eqref{eq:ODE}, with appropriate boundary conditions.

Since we wish to neglect the effect of boundary conditions in our analysis, we will solve \eqref{eq:ODE} on an unbounded domain with appropriate decay conditions as $|X|\to\infty$. We will study two different classes of gradient function in this work. First, we will consider symmetric gradient functions $g$, where we expect waves to be localised at the centre of the structure. In this case, the amplitude $f$ will be assumed to decay as $X\to\pm\infty$. Subsequently, we will study the more classical example of monotonic gradient functions $g$, where we expect the wave to propagate until it reaches a position of localisation. In this case, we assume that $f$ and its derivative decay as $X\to\infty$.

\section{Comparison of localisation for symmetric gradients} \label{sec:symmetric}

We begin by considering examples where the gradient function $g$ is symmetric. In which case, waves will be localised in the centre of the structure with decay in either direction. This can be thought of either as attaching two classical monotonic graded metamaterials back to back, or simply as an interesting wave localisation problem in its own right. Structures of this kind could be used as the starting point for building multi-dimensional waveguides that guide waves along specified paths. 

The class of symmetric gradients is a natural starting point as the subtlety of identifying the position of localisation is negated, since we expect waves to be localised at the centre. We will assume that the wave is at the cutoff frequency at the middle of the array, which equates to assuming that $g(0)=0$ and means we expect the wave amplitude to decay immediately in both directions. An additional advantage is that there are known explicit solutions to the Schr\"odinger equation \eqref{eq:ODE} for several symmetric gradient functions $g$, allowing us to validate our numerical approach.

Altogether, we will consider seven different gradient functions. These include the examples considered previously in the literature: linear \cite{deponti2020graded}, exponential \cite{rg_exponential, ammari2020mimicking}, cubic \cite{zhao2022graded} and cube root \cite{zhao2022graded}. We supplement this with three further examples, for which the Schr\"odinger equation \eqref{eq:ODE} has explicit analytic solutions. The seven gradient functions are shown in Figure~\ref{fig:symmetric}, in the first column of subplots. For the sake of calibrating the comparisons, the gradient functions are each normalised so that $g(0)=0$ and $g(\pm1)=1$, meaning that each gradient function has the same average gradient over the interval $[0,1]$.

\subsection{Numerical results}

\begin{figure}
    \centering
    \includegraphics[width=0.8\linewidth]{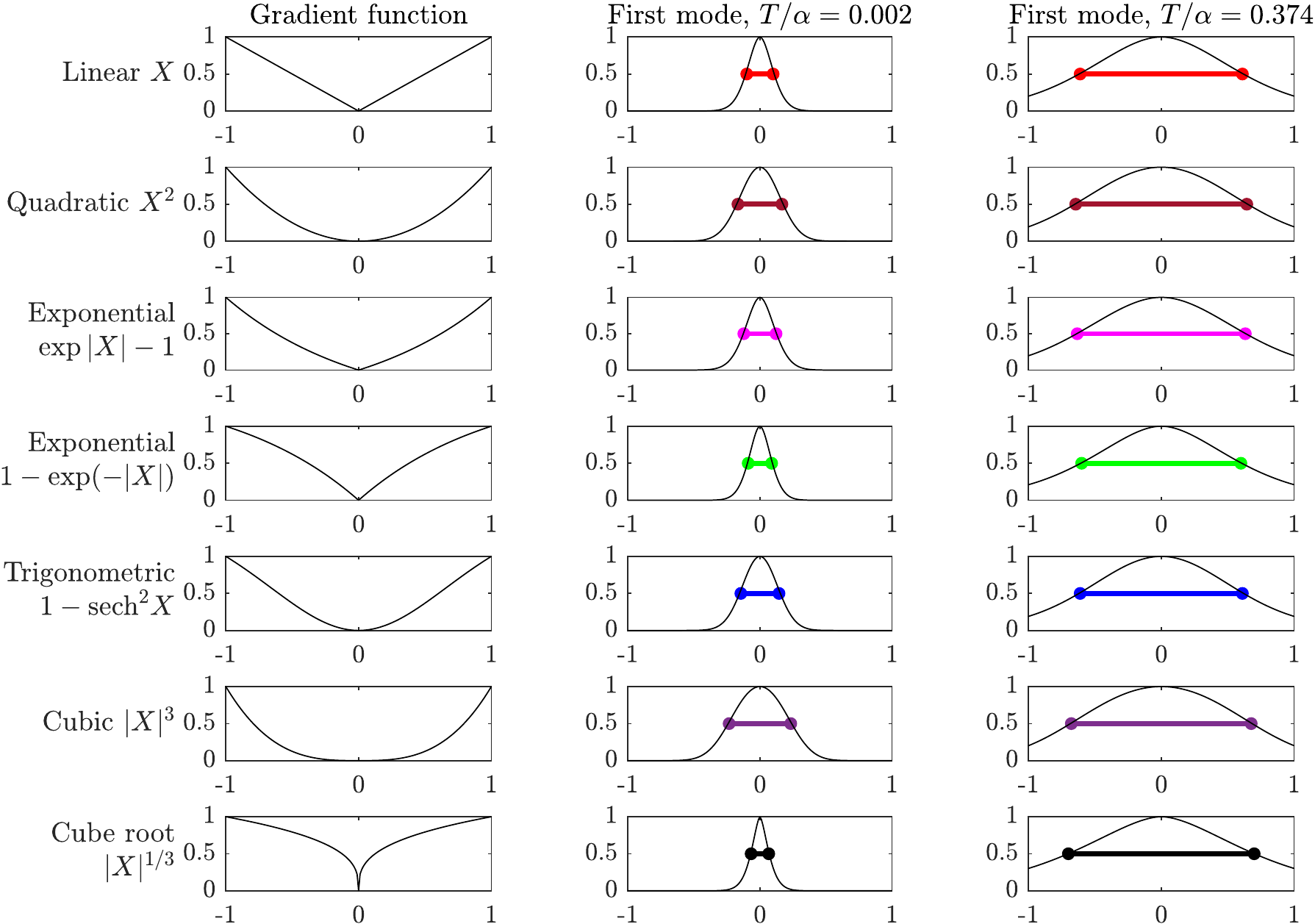}

    \vspace{0.7cm}

    \includegraphics[width=0.8\linewidth]{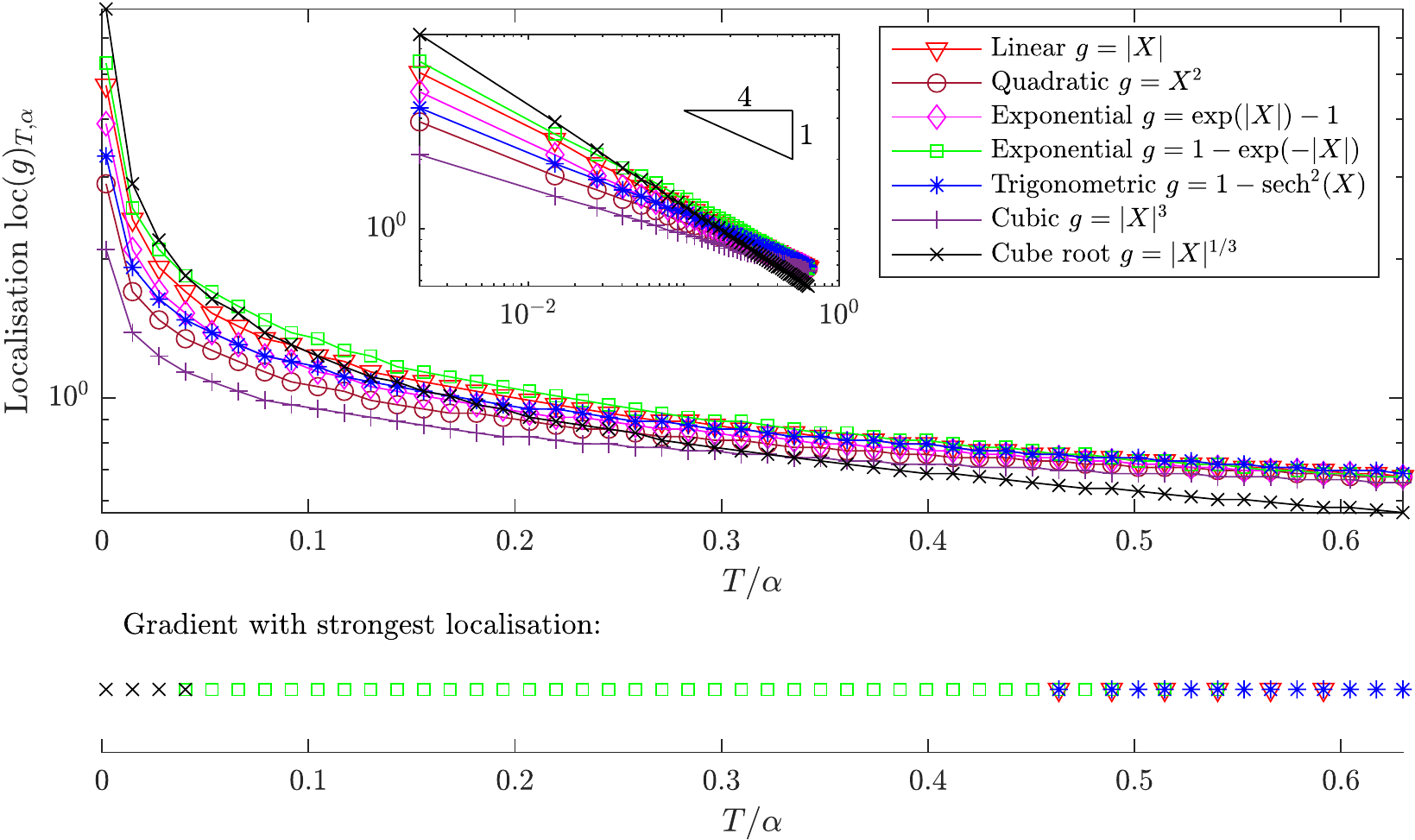}
    
    \caption{Symmetric gradient functions give symmetric eigenmodes whose localisation can be quantified using the full width at half maximum (FWHM). In the upper displays, we plot the first eigenmode of \eqref{eq:ODE} for seven different symmetric gradient functions (and two different values of $T/\alpha$). On each mode, the length corresponding to the FWHM is shown. Below this, the localisation (given by the reciprocal of the FWHM) is plotted as a function of $T/\alpha$ for each symmetric gradient function (the inset plot shows the same data on a log--log plot). Finally, the lower display shows the gradient with the most localised first eigenmode at each $T/\alpha$ (with multiple symbols shown in the event of a tie).}
    \label{fig:symmetric}
\end{figure}

Numerical solutions to the eigenvalue problem \eqref{eq:ODE} can be computed using a spectral method with periodic spectral differentiation \cite{trefethen2000spectral}. In Figure~\ref{fig:symmetric}, examples of the localised mode profiles are shown. In each row of subplots, the gradient function is plotted followed by the first eigenmode for two different values of $T/\alpha$. The full width at half maximum (FWHM) is indicated on each eigenmode. Subsequently, the localisation \eqref{def:loc} can be studied as a function of $T/\alpha$, which is shown in the lower part of \Cref{fig:symmetric}. The same data points are repeated on a log--log scale inset, where it is apparent that each line is approximately a straight line. Below the main plot, we show the gradient that gave the strongest localisation. It is notable that multiple symbols are shown for some values of $T/\alpha$, which corresponds to the modes being equally localised (at the resolution of the numerical scheme).

A striking conclusion to draw from \Cref{fig:symmetric} is that the relative strengths of localisation of the different gradient functions changes as $T/\alpha$ varies. The straight lines in the inset log--log plot have different slopes for different gradient functions, meaning the range of $T/\alpha$ is divided into three regimes, which is clear from looking at the bottom display in \Cref{fig:symmetric}. For very small values of $T/\alpha$, the cube root gradient gives the strongest localisation. As $T/\alpha$ increases, we then enter an intermediate regime where the concave exponential gradient performs best. Finally, for large values of $T/\alpha$, we see that the trigonometric and linear gradients give the strongest localisation. This suggests that, for the case of symmetric gradient functions, the gradient which gives the strongest localisation will depend on the specific physical setting and choice of metamaterial (as this determines the values of $T$ and $\alpha$).

\subsection{Analytic results}

The Schr\"odinger equation \eqref{eq:ODE} has known analytic solutions for some symmetric potentials. This means the numerical results of \Cref{fig:symmetric} can be partially verified analytically. The first example which admits an explicit solution is $g(X)=1-\sech^2X$. The localised eigenmode, and associated homogenised envelope function $f$, was shown in Figure~\ref{fig:sechmode}. In this case, the (normalised) solution $f$ of the ODE \eqref{eq:ODE} is given by \cite{infeld1951factorization}
\begin{equation} \label{eq:fsech}
    f(X)=\cosh^{1/2-\gamma}X,
\end{equation}
where $\gamma=\sqrt{1/4+\alpha/T}$. Additionally, the case $g(X)=X^2$ admits the well-known solution
\begin{equation} \label{eq:fquad}
    f(X)=\exp\bigg(-\frac{1}{2}\sqrt{\frac{\alpha}{T}}X^2\bigg).
\end{equation}
Finally, the Schr\"odinger equation \eqref{eq:ODE} can be reduced to the Bessel equation under appropriate transformations. For example, if $g(X)=1-\exp(-|X|)$ then this approach can be used to see that \cite{sasaki2016one}
\begin{equation} \label{eq:fexp}
    f(X)=J_\nu\bigg(2\sqrt{\frac{\alpha}{T}}\exp\Big(-\tfrac{|X|}{2}\Big)\bigg),
\end{equation}
where $J_\nu$ is the standard Bessel function of the first kind and the order $\nu$ is given by $\nu=2\sqrt{\alpha/T}(\Omega_2^2+1)$. The order $\nu$ can also be found directly, without knowing the value of the eigenvalue $\Omega_2^2$, by solving the implicit equation $J_{\nu-1}(2\sqrt{\alpha/T})-\nu/2\sqrt{T/\alpha}J_{\nu}(2\sqrt{\alpha/T})=0$ \cite{sasaki2016one}.

Given these explicit solutions \eqref{eq:fsech}--\eqref{eq:fexp}, it is possible to compute the localisation \eqref{def:loc} explicitly. We have that
\begin{equation}
    \loc(1-\sech^2X)_{T,\alpha}=\left( 2\cosh^{-1} \big( 2^{\frac{2}{2\gamma-1}} \big) \right)^{-1},
\end{equation}
where $\cosh^{-1}:[1,\infty)\to[0,\infty)$ is the inverse hyperbolic cosine function. Additionally, the convenient form of the solution \eqref{eq:fquad} for the quadratic potential leads to the expression
\begin{equation}
    \loc(X^2)_{T,\alpha}=\left( 2\sqrt{2\sqrt{\tfrac{T}{\alpha}}\log2} \right)^{-1}.
\end{equation}
This is particularly informative, as it shows that $\loc(X^2)_{T,\alpha}\propto(T/\alpha)^{-1/4}$, explaining the shape of the curve observed in Figure~\ref{fig:symmetric}. The rate triangle on the inset log--log plot in Figure~\ref{fig:symmetric} shows the expected $\loc_{T,\alpha}\propto(T/\alpha)^{-1/4}$ rate, for comparison. A similar formula can be obtained from \eqref{eq:fexp}, however it relies on inverse Bessel functions so is difficult to state concisely, hence we omit it for brevity.

\section{Comparison of localisation for monotonic gradients} \label{sec:monotonic}

We now consider the classical case of gradient functions $g$ that are monotonically increasing. This case is widely used in energy harvesting and machine hearing settings, as it allows waves to couple with and propagate through the metamaterial up until the point at which they are localised (which depends on their frequency). Our definition of the (modified) full width at half maximum \eqref{eq:fwhm_defn} is designed so that it only measures the width of the largest peak, which will be the final peak before the mode's amplitude decays quickly. An important subtlety for monotonic gradient functions is that, in contrast to the symmetric gradients studied in the previous setting, it is not immediately clear \emph{where} the wave will be localised. Indeed, characterising the position at which it is localised will be a key step in our analysis, below.

We consider seven different gradient functions, which are the monotonic versions of the symmetric gradient functions studied in the previous section. In order to make for a fair comparison between the different gradients, we normalise the average gradient to be 1 across the interval $[-1,1]$, achieved by setting $g(-1)=-1$ and $g(1)=1$. For physical reality, so we don't have arbitrarily large physical structures, the gradient is truncated such that $g(X)=-1$ for $X<-1$ and $g(X)=1$ for $X>1$. 

\subsection{Numerical results}

\begin{figure}
    \centering
    \includegraphics[width=0.8\linewidth]{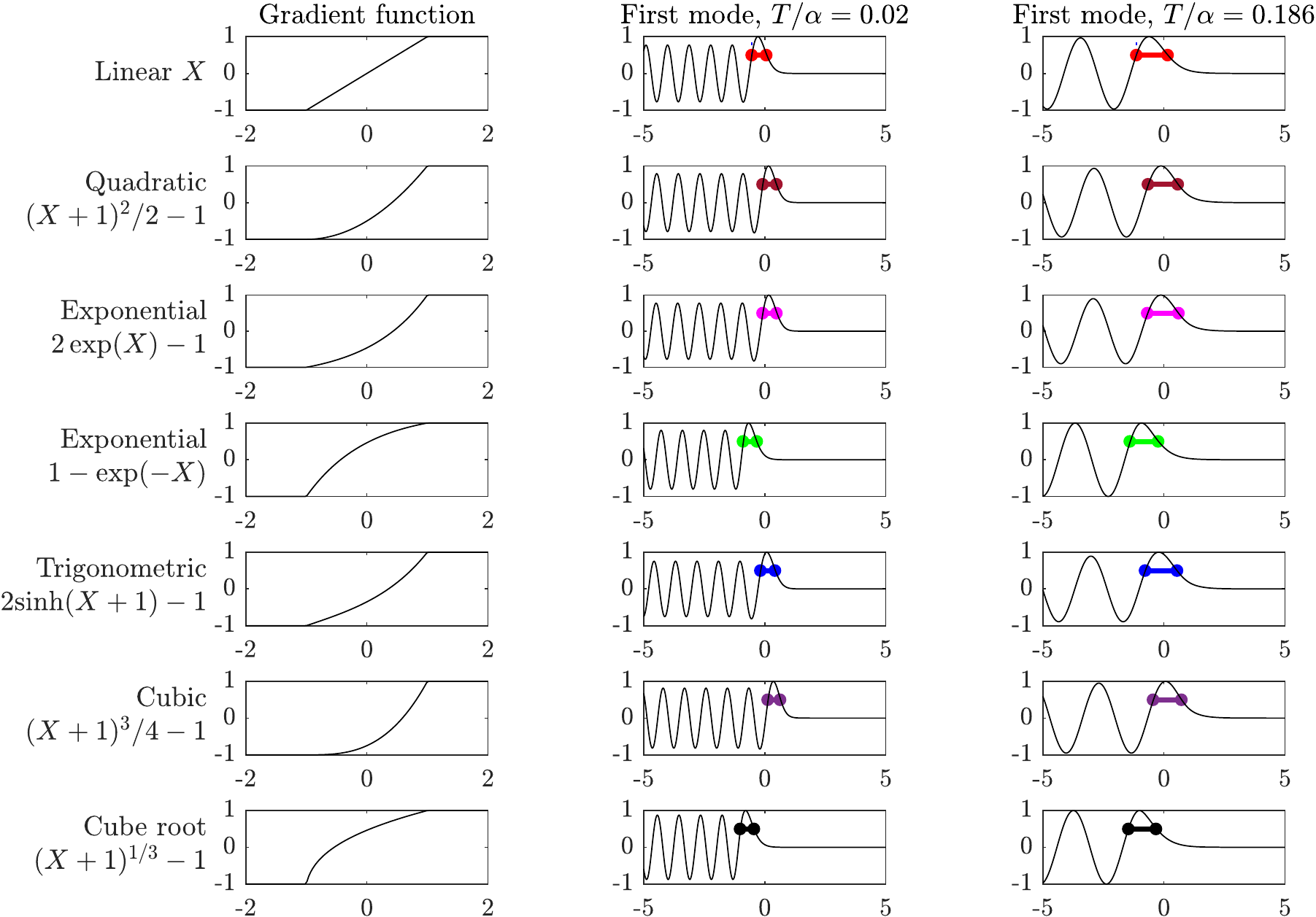}

    \vspace{0.7cm}

    \includegraphics[width=0.8\linewidth]{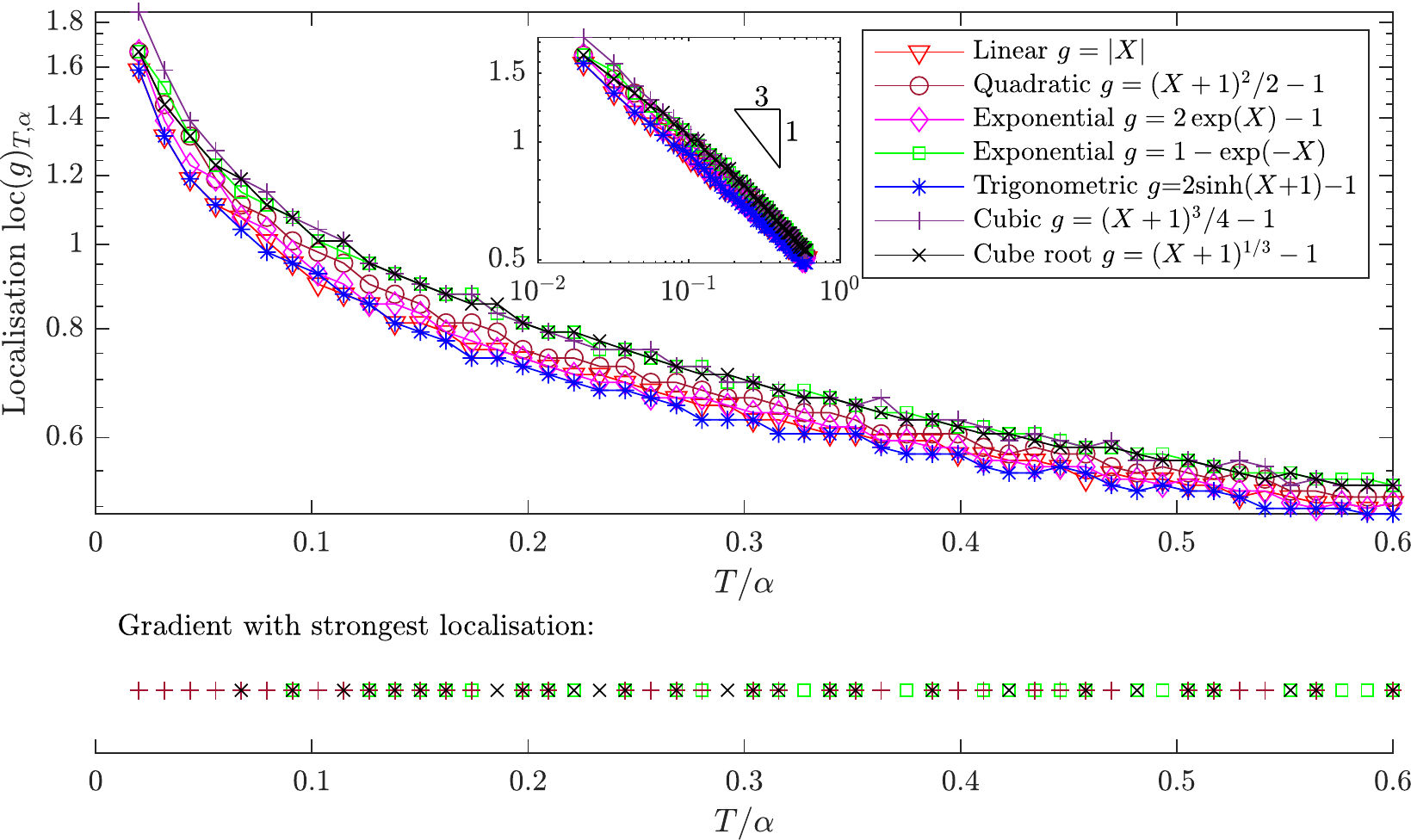}
    
    \caption{Monotonic gradient functions give eigenmodes that transition from bring a propagating mode to a decaying mode. Their localisation can be quantified using the full width at half maximum (FWHM) of the last peak. In the upper displays, we plot the first eigenmode of \eqref{eq:ODE} for seven different monotonic gradient functions (and two different values of $T/\alpha$). On each mode, the length corresponding to the FWHM is shown. Below this, the localisation (given by the reciprocal of the FWHM) of the first mode is plotted for different values of $T/\alpha$ for each gradient function. Finally, the lower display shows the gradient with the most localised first eigenmode at each $T/\alpha$ (with multiple symbols shown in the event of a tie).}
    \label{fig:monotonic}
\end{figure}

The numerical results for monotonic gradients are shown in \Cref{fig:monotonic}. In this case, it should be noted immediately that the differences between the different gradient functions are much smaller than for the symmetric gradients. Whereas it was possible to judge by eye which modes in Figure~\ref{fig:symmetric} were most or least strongly localised, the differences are much smaller in \Cref{fig:monotonic}. Another difference from the symmetric case is that there are no apparent transitions between the relative localisation of different gradients. The lines in the log--log plot in \Cref{fig:monotonic} are (approximately) parallel and there appears to be a consistent hierarchy of the localisation due to each gradient. This hierarchy is that the cubic, cube root and concave exponential gradients produce the strongest localisation (with minimal differences between them), followed by the quadratic gradient and subsequently by the convex exponential. Finally, the trigonometric and linear gradients give the weakest localisation (again, with differences at the resolution of our numerical scheme).

These results, firstly, validate the conclusions of \cite{rg_exponential, rg_linear} that exponential gradients lead to stronger localisation than linear gradients (it turns out, this is irrespective of whether the exponential is convex or concave). Additionally, they also confirm the conclusions of \cite{zhao2022graded} that cube root gradients similarly outperform linear gradients. That these conclusions hold consistently across a range of values of $T/\alpha$ suggests that these conclusions will hold for a broad range of physical systems (so long as they fall into the multi-scale framework of the high-frequency homogenisation methodology).

\subsection{Analytic results}

In the case of monotonic gradient functions, analytic solutions to the Schr\"odinger equation \eqref{eq:ODE} are harder to come by. The notable exception to this is the case of a linear gradient, for which \eqref{eq:ODE} is a transformed version of the famous Airy equation \cite{bender1999advanced}. We can see this by making the change of variables 
\begin{equation}
    Y= \left(\frac{T}{\alpha}\right)^{2/3} \frac{\alpha X-\Omega_2^2}{T},
\end{equation}
which transforms \eqref{eq:ODE} into the Airy equation
\begin{equation} \label{airy}
    f''(Y)=Yf(Y).
\end{equation}
Since we are interested in solutions for which $f(X)\to0$ as $X\to\infty$, this determines the correct solution of \eqref{airy} and we find that the solution of \eqref{eq:ODE} with $g(X)=X$ is given by
\begin{equation} \label{eq:airysoln}
    f(x)=\Ai\bigg[ \left(\frac{T}{\alpha}\right)^{2/3} \frac{\alpha X-\Omega_2^2}{T} \bigg],
\end{equation}
where $\Ai[\, \cdot \, ]$ is the Airy function of the first kind \cite{bender1999advanced}. 

For non-linear monotonic gradient functions we will use approximation strategies in order to make useful analytic comparisons\footnote{In reality, the solution to the Schr\"odinger equation can be expressed in terms of special functions for many different non-linear monotonic gradient functions. For example, parabolic cylinder functions can be used for quadratic gradients and modified Bessel functions of the first kind can be used for cubic gradients. However, such solutions are of limited use for drawing comparisons and it will be much more insightful to approximate each solution locally by an Airy function.}. Since our aim is to compare the extent of localisation, our approach will be to linearise $g$ around the point at which it is localised, and then formulate an approximate solution in terms of Airy functions. Before doing this, we can use a \emph{WKB approximation} to characterise the overall properties of the solution. This approach, named after Wentzel, Kramers and Brillouin, is a popular method for approximating the solution to the Schr\"odinger equation in the case that the gradient function is slowly varying \cite{bender1999advanced}. The idea is that, if the gradient varies slowly, the solution $f(X)$ will be similar to the oscillating function $\exp(\I kX)$ that would solve the equation if the gradient $g$ was constant. In particular, the WKB ansatz is 
\begin{equation} \label{eq:WKB}
    f(X) = A(X) e^{\I \Phi(X)},
\end{equation}
for some real-valued functions $A$ and $\Phi$ that are to be found. Substituting \eqref{eq:WKB} into \eqref{eq:ODE} and equating real and imaginary parts gives the equations $(A^2\Phi')'=0$ and $A''=A((\Phi')^2-k^2)$, where $k=k(X)$ is a spatially varying local wavenumber, given by
\begin{equation}
    k(X) = \sqrt{\frac{\Omega_2^2-\alpha g(X)}{T}}.
\end{equation}
The first of these two equations implies that $A=C/\sqrt{\Phi'}$ for some constant $C$. The second of these can be simplified by assuming that $A''/A$ is much smaller than both $(\Phi')^2$ and $k^2$. In which case, we have that $\Phi'=\pm k$ and we reach the solution
\begin{equation} \label{eq:WKBsoln}
    f(X) \approx \frac{1}{\sqrt{k(X)}} \exp\bigg( \pm\I \int^X k(s) \de s \bigg).
\end{equation}
This approximate solution can be used to understand the overall behaviour of $f$. When $\alpha g(X)<\Omega_2^2$ the wavenumber $k(X)$ is real-valued meaning that the approximate solution \eqref{eq:WKBsoln} for $f$ corresponds to an oscillatory propagating wave. Conversely, when $\alpha g(X)>\Omega_2^2$ the wavenumber $k(X)$ is purely imaginary meaning that the solution has an exponentially decaying profile. This matches the behaviour we see in Figure~\ref{fig:monotonic}.

It is clear that the approximate WKB solution \eqref{eq:WKBsoln} is not valid near to the point where $\alpha g(X)=\Omega_2^2$, as the wavenumber $k(X)$ vanishes there. In this region, we will linearise the gradient about this point, which reduces the Schr\"odinger equation \eqref{eq:ODE} to an Airy equation \eqref{airy}. In particular, we have that 
\begin{equation} \label{airy_approx}
    f''(X)=\frac{\alpha g'(X_0)}{T}(X-X_0)f(X)+O\big((X-X_0)^2\big),
\end{equation}
as $X-X_0\to0$, where $X_0$ is the point such that $\alpha g(X_0)=\Omega_2^2$. This leads to the approximate solution
\begin{equation} \label{eq:Airyapprox}
    f(X)=\Ai\bigg[ \left(\frac{\alpha g'(X_0)}{T}\right)^{1/3} (X-X_0)\bigg]+O\big((X-X_0)^2\big).
\end{equation}
Based on this analysis, it is possible to form an approximate solution for $f$ on the whole real line by stitching together the WKB solution \eqref{eq:WKBsoln} and the Airy solution \eqref{eq:Airyapprox} near to the singularity. This approach is similar to conventional boundary layer analyses \cite{bender1999advanced, verhulst2005methods} or the matched HFH-WKB method developed by \cite{schnitzer2017waves}. For our purposes, however, the expressions \eqref{eq:WKBsoln} and \eqref{eq:Airyapprox} are already sufficient.

As was the case for the symmetric gradients, we can use our analysis to estimate the localisation factor $\loc(g)_{T,\alpha}$. For the linear gradient, \eqref{eq:airysoln} shows us immediately that $\loc(X)_{T,\alpha}\propto(T/\alpha)^{-1/3}$. The approximate solution \eqref{eq:Airyapprox} suggests that this relationship will hold for other gradients also, which is confirmed by the inset log--log plot in Figure~\ref{fig:monotonic} and explains the observation that these lines all have the same slope.

\begin{figure}
    \centering
    \includegraphics[width=0.8\linewidth]{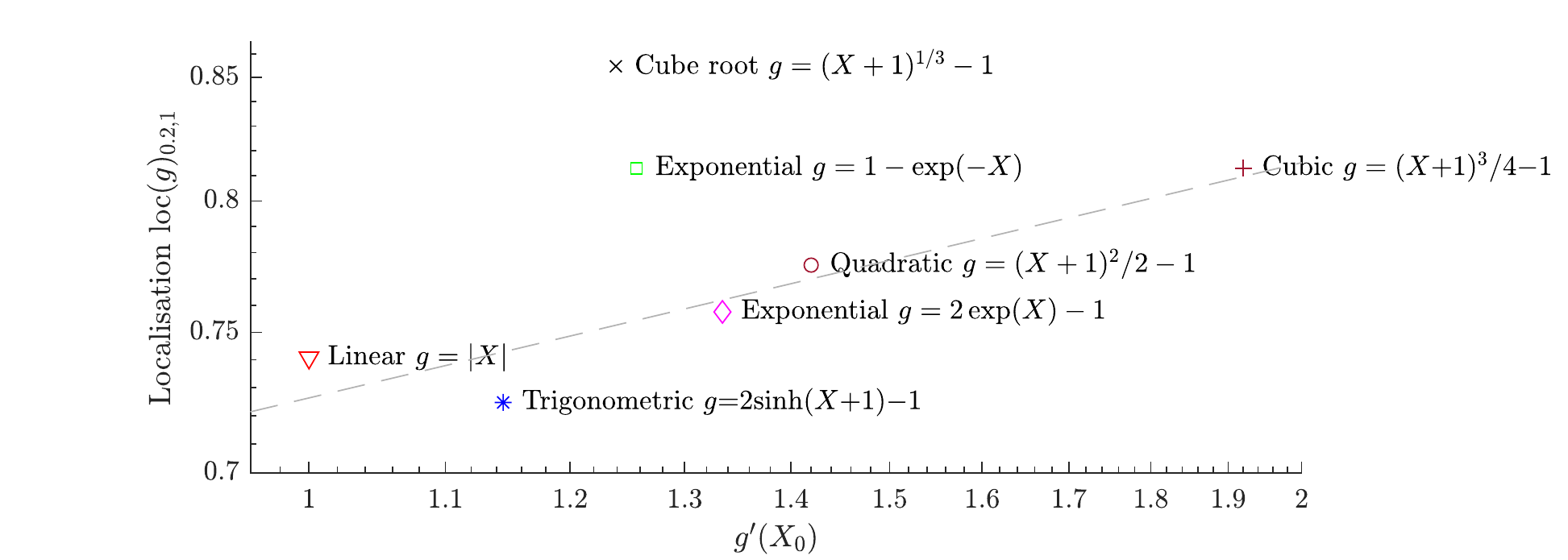}
    \caption{A comparison of the localisation $\mathrm{loc}(g)_{0.2,1}$ and the slope of the gradient function $g'(X_0)$, evaluated. Both axes use logarithmic scales. In this figure, we use $T=0.2$, $\alpha=1$ but note that the behaviour is the same for other parameter values. The dashed line is a straight line that has been fitted (in a least-squares sense) to the data points for the five convex gradient functions. }
    \label{fig:linearapprox}
\end{figure}

The approximation \eqref{eq:Airyapprox} reveals some insight into the significant role that the quantity $g'(X_0)$ plays in determining the strength of localisation. It suggests that the localisation can be maximised by maximising $g'(X_0)$. We can explore this hypothesis for the seven gradient functions considered in this work by comparing $\loc(g)_{T,\alpha}$ and $g'(X_0)$, which is shown in Figure~\ref{fig:linearapprox} for $T=0.2$ and $\alpha=1$ (the plot is qualitatively the same for other values of $T$ and $\alpha$). It is noticeable that the five convex gradient functions do indeed follow an approximate linear trend (on this log--log plot); the dashed line shown has been fitted to these five data points (in a least-squares sense) and has an $R$-squared value of $R^2=0.86$. This suggests that a viable strategy for maximising the localisation due to convex monotonic gradient functions is to maximise the slope of the gradient function $g$ at the point at which $\alpha g(X_0)=\Omega_2^2$. The cube root and concave exponential gradient functions do not appear to fit the trend that is otherwise present in Figure~\ref{fig:linearapprox}. Our interpretation of this is that it is a consequence of their concavity as it means the gradient function increases very quickly at the start of the graded region. This effectively leads to an impedance mismatch that localises the wave through reflection, which is a different mechanism to that which is employed by the gradient functions that sweep upwards more gradually.

\section{Concluding remarks}

High-frequency homogenisation provides a simple ordinary differential model that accurately describes wave localisation by a graded metamaterial. We have used this framework to compare the effect of the choice of gradient function in an efficient way, without the need for expensive wave scattering simulations. Additionally, the effective model used in this work means that our conclusions span several different physical settings and wave regimes (as we demonstrated in Section~\ref{sec:homogenisation}). Through this approach, we observed that the localisation due to symmetrically graded metamaterials varies depending on the effective parameters. This means that the symmetric gradient function that yields the strongest localisation will be different for different physical settings. For monotonically graded metamaterials the picture is slightly clearer, as there is a hierarchy of different gradient functions that is stable as the effective parameters change. Further, our analysis revealed that the localisation could be maximised (at least within the space of convex gradient functions) by maximising the slope of the gradient function at a specified transition point. However, it is important to note that, even in spite of this insight, the differences between some of the monotonic gradient functions are negligibly small.

The fact that no clear universal optimal choice of gradient function exists even in this simplified setting, where we have considered single frequency localisation within a homogenised model, provides important context for future studies. In particular, our results show the need to think carefully about the choice of objective function before applying optimisation or machine learning algorithms and that any such approach is likely to encounter a great number of local minima. Conversely, a more positive interpretation of these results is that graded metamaterials perform well for a wide range of gradient functions (and it doesn't seem to matter too much which gradient function is chosen). The fundamental mechanism behind the function of graded metamaterials (as depicted in \Cref{fig:sketches}) is sufficiently simple and powerful that it is (within reason) robust with respect to the choice of gradient.

% \todo[inline]{Why is this the right notion of localisation? - to make it coincide with some piezeelectric patch or a specific collection of receptor cells, to maximise resolution}

% \todo[inline]{Whether this is the correct or incorrect definition of localisation is something that \textit{could} be mentioned here, but it should probably be a discussion to be had in the conclusion section. I think we should be clear and specific, in that section, that if we care about single frequency maximal energy harvesting at a specific location(s), this is the measure we care about. There could be a short paragraph that discusses other measures, but we need to be careful so as to not undermine all of this analysis.}

% Suggested strategy is to maximise $g'(X_0)$, at least for convex gradient functions 

% Implications for signal processing and machine hearing \cite{ammari2023asymptotic, lyon2017human}.

% We have just done single frequency, could do broadband in future

\section*{Acknowledgements}
The authors would like to thank Richard Craster and Marie Touboul for helpful advice and suggestions. 

\section*{Funding}
The work of B.D. was funded by EPSRC through a fellowship with grant number EP/X027422/1. H.J.P. is grateful for EPSRC funding provided through the Centre for Doctoral Training ‘Fluid Dynamics Across Scales’ Programme (grant no. EP/L016230/1. R.V).

\section*{Data Accessibility}
The software used to perform the numerical simulations presented in this work is available for download at \url{https://doi.org/10.5281/zenodo.7864009}.

\section*{Conflicts of interest}
The authors declare that they have no conflicts of interest.

% Bibliography
\bibliographystyle{unsrt}
\bibliography{references}

\begin{thebibliography}{10}

\bibitem{joannopoulos1995photonic}
J.~D. Joannopoulos, R.~D. Meade, and J.~N Winn.
\newblock {\em Photonic Crystals, Molding the Flow of Light}.
\newblock Princeton University Press, Princeton, NJ, 1995.

\bibitem{smith2004metamaterials}
D.~R. Smith, J.~B. Pendry, and M.~C.~K. Wiltshire.
\newblock Metamaterials and negative refractive index.
\newblock {\em Science}, 305(5685):788--792, 2004.

\bibitem{miltonbook}
G.~W. Milton.
\newblock {\em The theory of composites}.
\newblock Cambridge University Press, Cambridge, UK, 2002.

\bibitem{khanikaev2013photonic}
A.~B. Khanikaev, S.~H. Mousavi, W.-K. Tse, M.~Kargarian, A.~H. MacDonald, and
  G.~Shvets.
\newblock Photonic topological insulators.
\newblock {\em Nat. Mater.}, 12(3):233--239, 2013.

\bibitem{rechtsman2013photonic}
M.~C. Rechtsman, J.~M. Zeuner, Y.~Plotnik, Y.~Lumer, D.~Podolsky, F.~Dreisow,
  S.~Nolte, M.~Segev, and A.~Szameit.
\newblock Photonic {Floquet} topological insulators.
\newblock {\em Nature}, 496(7444):196--200, 2013.

\bibitem{pendry2000negative}
J.~B. Pendry.
\newblock Negative refraction makes a perfect lens.
\newblock {\em Phys. Rev. Lett.}, 85(18):3966, 2000.

\bibitem{milton2006cloaking}
G.~W. Milton and N.-A.~P. Nicorovici.
\newblock On the cloaking effects associated with anomalous localized
  resonance.
\newblock {\em Proc. R. Soc. A}, 462(2074):3027--3059, 2006.

\bibitem{kadic20193d}
M.~Kadic, G.~W. Milton, M.~van Hecke, and M.~Wegener.
\newblock 3d metamaterials.
\newblock {\em Nat. Rev. Phys.}, 1(3):198--210, 2019.

\bibitem{tsakmakidis2007trapped}
K.~L. Tsakmakidis, A.~D. Boardman, and O.~Hess.
\newblock {‘Trapped} rainbow’ storage of light in metamaterials.
\newblock {\em Nature}, 450(7168):397--401, 2007.

\bibitem{zhu2013acoustic}
J.~Zhu, Y.~Chen, X.~Zhu, F.~J. Garcia-Vidal, X.~Yin, W.~Zhang, and X.~Zhang.
\newblock Acoustic rainbow trapping.
\newblock {\em Sci. Rep.}, 3(1):1--6, 2013.

\bibitem{arreola2019experimental}
A.~Arreola-Lucas, G.~Baez, F.~Cervera, A.~Climente, R.~A.
  M{\'e}ndez-S{\'a}nchez, and J.~S{\'a}nchez-Dehesa.
\newblock Experimental evidence of rainbow trapping and {Bloch} oscillations of
  torsional waves in chirped metallic beams.
\newblock {\em Sci. Rep.}, 9(1):1--13, 2019.

\bibitem{skelton2018multi}
E.~A. Skelton, R.~V. Craster, A.~Colombi, and D.~J. Colquitt.
\newblock The multi-physics metawedge: graded arrays on fluid-loaded elastic
  plates and the mechanical analogues of rainbow trapping and mode conversion.
\newblock {\em New J. Phys.}, 20(5):053017, 2018.

\bibitem{colombi2016seismic}
A.~Colombi, D.~Colquitt, P.~Roux, S.~Guenneau, and R.~V. Craster.
\newblock A seismic metamaterial: The resonant metawedge.
\newblock {\em Sci. Rep.}, 6(1):27717, 2016.

\bibitem{bennetts2019low}
L.~G. Bennetts, M.~A. Peter, and R.~V. Craster.
\newblock Low-frequency wave-energy amplification in graded two-dimensional
  resonator arrays.
\newblock {\em Phil. Trans. R. Soc. A}, 377(2156):20190104, 2019.

\bibitem{chaplain2020delineating}
G.~J. Chaplain, D.~Pajer, J.~M. De~Ponti, and R.~V. Craster.
\newblock Delineating rainbow reflection and trapping with applications for
  energy harvesting.
\newblock {\em New J. Phys.}, 22(6):063024, 2020.

\bibitem{zhao2022graded}
B.~Zhao, H.~R. Thomsen, J.~M. De~Ponti, E.~Riva, B.~Van~Damme, A.~Bergamini,
  E.~Chatzi, and A.~Colombi.
\newblock A graded metamaterial for broadband and high-capability piezoelectric
  energy harvesting.
\newblock {\em Energy Convers. Manage.}, 269:116056, 2022.

\bibitem{deponti2020graded}
J.~M. De~Ponti, A.~Colombi, R.~Ardito, F.~Braghin, A.~Corigliano, and R.~V.
  Craster.
\newblock Graded elastic metasurface for enhanced energy harvesting.
\newblock {\em New J. Phys.}, 22(1):013013, 2020.

\bibitem{ammari2020mimicking}
H.~Ammari and B.~Davies.
\newblock Mimicking the active cochlea with a fluid-coupled array of
  subwavelength {Hopf} resonators.
\newblock {\em Proc. R. Soc. A}, 476(2234):20190870, 2020.

\bibitem{marrocchio2021waves}
R.~Marrocchio, A.~Karlos, and S.~Elliott.
\newblock Waves in the cochlea and in acoustic rainbow sensors.
\newblock {\em Wave Motion}, 106:102808, 2021.

\bibitem{rupin2019mimicking}
M.~Rupin, G.~Lerosey, J.~de~Rosny, and F.~Lemoult.
\newblock Mimicking the cochlea with an active acoustic metamaterial.
\newblock {\em New J. Phys.}, 21(9):093012, 2019.

\bibitem{antonakakis2014asymptotic}
T.~Antonakakis, R.~V. Craster, S.~Guenneau, and E.~A. Skelton.
\newblock An asymptotic theory for waves guided by diffraction gratings or
  along microstructured surfaces.
\newblock {\em Proc. R. Soc. A}, 470(2161):20130467, 2014.

\bibitem{chaplain2019rayleigh}
G.J. Chaplain, M.P. Makwana, and R.V. Craster.
\newblock Rayleigh--bloch, topological edge and interface waves for structured
  elastic plates.
\newblock {\em Wave Motion}, 86:162--174, 2019.

\bibitem{ammari2019fully}
H.~Ammari and B.~Davies.
\newblock A fully coupled subwavelength resonance approach to filtering
  auditory signals.
\newblock {\em Proc. R. Soc. A}, 475(2228):20190049, 2019.

\bibitem{rg_linear}
V.~Romero-García, R.~Picó, A.~Cebrecos, V.~J. Sánchez-Morcillo, and
  K.~Staliunas.
\newblock Enhancement of sound in chirped sonic crystals.
\newblock {\em Appl. Phys. Lett.}, 102(9):091906, 2013.

\bibitem{rg_exponential}
A.~Cebrecos, R.~Picó, V.~J. Sánchez-Morcillo, K.~Staliunas,
  V.~Romero-García, and L.~M. Garcia-Raffi.
\newblock Enhancement of sound by soft reflections in exponentially chirped
  crystals.
\newblock {\em AIP Adv.}, 4(12):124402, 2014.

\bibitem{alshaqaq2020graded}
M.~Alshaqaq and A.~Erturk.
\newblock Graded multifunctional piezoelectric metastructures for wideband
  vibration attenuation and energy harvesting.
\newblock {\em Smart Mater. Struct.}, 30(1):015029, 2020.

\bibitem{jimenez2017rainbow}
N.~Jim{\'e}nez, V.~Romero-Garc{\'\i}a, V.~Pagneux, and J.-P. Groby.
\newblock Rainbow-trapping absorbers: Broadband, perfect and asymmetric sound
  absorption by subwavelength panels for transmission problems.
\newblock {\em Sci. Rep.}, 7(1):13595, 2017.

\bibitem{wilks2022rainbow}
B.~Wilks, F.~Montiel, and S.~Wakes.
\newblock Rainbow reflection and broadband energy absorption of water waves by
  graded arrays of vertical barriers.
\newblock {\em J. Fluid Mech.}, 941:A26, 2022.

\bibitem{rosafalco2022optimised}
L.~Rosafalco, J.~M. De~Ponti, L.~Iorio, R.~Ardito, and A.~Corigliano.
\newblock Optimised graded metamaterials for mechanical energy confinement and
  amplification via reinforcement learning.
\newblock {\em Eur. J. Mech. A Solids}, 99:104947, 2023.

\bibitem{davies2022robustness}
B.~Davies and L.~Herren.
\newblock Robustness of subwavelength devices: a case study of cochlea-inspired
  rainbow sensors.
\newblock {\em Proc. R. Soc. A}, 478(2262):20210765, 2022.

\bibitem{chaplain2020topological}
G.~J. Chaplain, J.~M. De~Ponti, G.~Aguzzi, A.~Colombi, and R.~V. Craster.
\newblock Topological rainbow trapping for elastic energy harvesting in graded
  {Su-Schrieffer-Heeger} systems.
\newblock {\em Phys. Rev. Appl.}, 14(5):054035, 2020.

\bibitem{davies2023graded}
B.~Davies, G.~J. Chaplain, T.~A. Starkey, and R.~V. Craster.
\newblock Graded quasiperiodic metamaterials perform fractal rainbow trapping.
\newblock {\em arXiv preprint arXiv:2305.10520}, 2023.

\bibitem{craster2010high}
R.~V. Craster, J.~Kaplunov, and A.~V. Pichugin.
\newblock High-frequency homogenization for periodic media.
\newblock {\em Proc. R. Soc. A}, 466(2120):2341--2362, 2010.

\bibitem{allaire1992homogenization}
G.~Allaire.
\newblock Homogenization and two-scale convergence.
\newblock {\em SIAM J. Math. Anal.}, 23(6):1482--1518, 1992.

\bibitem{bensoussan2011asymptotic}
A.~Bensoussan, J.-L. Lions, and G.~Papanicolaou.
\newblock {\em Asymptotic Analysis for Periodic Structures}, volume 374.
\newblock American Mathematical Soc., 2011.

\bibitem{craster2022asymptotic}
R.~V. Craster and B.~Davies.
\newblock Asymptotic characterisation of localized defect modes:
  {Su-Schrieffer-Heeger} and related models.
\newblock {\em Multiscale Model. Simul.}, 21(3):827--848, 2023.

\bibitem{guzina2019rational}
B.~B. Guzina, S.~Meng, and O.~Oudghiri-Idrissi.
\newblock A rational framework for dynamic homogenization at finite wavelengths
  and frequencies.
\newblock {\em Proc. R. Soc. A}, 475(2223):20180547, 2019.

\bibitem{meng2022convergent}
S.~Meng, O.~Oudghiri-Idrissi, and B.~B. Guzina.
\newblock A convergent low-wavenumber, high-frequency homogenization of the
  wave equation in periodic media with a source term.
\newblock {\em Appl. Anal.}, 101(18):6451--6484, 2022.

\bibitem{touboul2023dispersive}
M.~Touboul, B.~Vial, R.~C. Assier, S.~Guenneau, and R.~V. Craster.
\newblock High-frequency homogenization for dispersive media.
\newblock {\em arXiv preprint}, 2023.

\bibitem{assier2020high}
R.~C. Assier, M.~Touboul, B.~Lombard, and C.~Bellis.
\newblock High-frequency homogenization in periodic media with imperfect
  interfaces.
\newblock {\em Proc. R. Soc. A}, 476(2244):20200402, 2020.

\bibitem{kronig1931quantum}
R.~de~L. Kronig and W.~G. Penney.
\newblock Quantum mechanics of electrons in crystal lattices.
\newblock {\em Proc. R. Soc. Lond. A}, 130(814):499--513, 1931.

\bibitem{trefethen2000spectral}
L.~N. Trefethen.
\newblock {\em Spectral Methods in {MATLAB}}.
\newblock SIAM, Philadelphia, 2000.

\bibitem{infeld1951factorization}
L.~Infeld and T.~E. Hull.
\newblock The factorization method.
\newblock {\em Rev. Mod. Phys.}, 23(1):21, 1951.

\bibitem{sasaki2016one}
R.~Sasaki and M.~Znojil.
\newblock One-dimensional {Schr{\"o}dinger} equation with non-analytic
  potential {V}(x)=-exp(-$|$x$|$) and its exact {Bessel}-function solvability.
\newblock {\em J. Phys. A: Math. Theor.}, 49(44):445303, 2016.

\bibitem{bender1999advanced}
C.~M. Bender and S.~A. Orszag.
\newblock {\em Advanced Mathematical Methods for Scientists and Engineers I:
  Asymptotic Methods and Perturbation Theory}.
\newblock Springer, 1999.

\bibitem{verhulst2005methods}
F.~Verhulst.
\newblock {\em Methods and Applications of Singular Perturbations: Boundary
  Layers and Multiple Timescale Dynamics}.
\newblock Springer, 2005.

\bibitem{schnitzer2017waves}
Ory Schnitzer.
\newblock Waves in slowly varying band-gap media.
\newblock {\em SIAM J. Appl. Math.}, 77(4):1516--1535, 2017.

\end{thebibliography}
\end{document}